\documentclass[preprint,superscriptaddress]{revtex4}
\usepackage{graphicx}
\usepackage{array}
\usepackage{amsfonts}
\usepackage{amssymb,amsmath,multirow,rotate}
\usepackage{mathrsfs}
\usepackage{booktabs}
\usepackage{threeparttable}
\usepackage{multirow}
\usepackage{epsfig}
\usepackage{threeparttable}
\usepackage{chngpage}
\usepackage{latexsym}
\usepackage{dcolumn}
\usepackage{graphics,graphicx,fleqn,epic,eepic}
\usepackage{times}
\usepackage{verbatim}
\usepackage{subfigure}
\usepackage{color}
\usepackage{float}
\usepackage{bm}

\definecolor{red}{rgb}{1,0,0}

\definecolor{blue}{rgb}{0,0,1}

\begin{document}

\title{Transition to turbulence in ferrofluids}

\author{Sebastian Altmeyer} 
\email{sebastian_altmeyer@t-online.de}
\affiliation{Institute of Science and Technology Austria 
(IST Austria), 3400 Klosterneuburg, Austria}

\author{Younghae Do}
\email{yhdo@knu.ac.kr}
\affiliation{Department of Mathematics,
KNU-Center for Nonlinear Dynamics, 
Kyungpook National University, Daegu, 702-701, South Korea}

\author{Ying-Cheng Lai}
\affiliation{School of Electrical, Computer and Energy Engineering, 
Arizona State University, Tempe, Arizona, 85287, USA}

\begin{abstract}
It is known that in classical fluids 
turbulence typically occurs at high Reynolds numbers. 
But can turbulence occur at low Reynolds numbers? 
Here we investigate the transition to turbulence in the classic Taylor-Couette system 
in which the rotating fluids are manufactured ferrofluids with 
magnetized nanoparticles embedded in liquid carriers. 
We find that, in the presence of a magnetic field 
turbulence can occur at Reynolds numbers 
that are at least one order of magnitude smaller than those in conventional fluids. 
This is established by extensive computational ferrohydrodynamics
through a detailed bifurcation analysis and 
characterization of behaviors of physical quantities 
such as the energy, the wave number, and the angular momentum through the bifurcations. 
A striking finding is that, as the magnetic field is increased, 
the onset of turbulence can be determined accurately and reliably. 
Our results imply that 
experimental investigation of turbulence can be greatly facilitated by using ferrofluids, 
opening up a new avenue to probe into the fundamentals of turbulence 
and the challenging problem of turbulence control. 

\end{abstract}
\date{\today}
\maketitle

Turbulence have been subject to intensive studies as a central issue in the modern science.
In classical Newtonian fluid, it is well known that 
the occurrence of turbulence depends on the Reynolds number~\cite{GS:1975,Frisch:book},
and typically in high Reynolds number.
Here the Reynolds number ($Re$) is a dimensionless quantity defined as $Re \equiv l u/\nu$, 
where $l$ and $u$ are the characteristic length scale and a typical velocity of the flow 
in a particular geometry, respectively, and $\nu$ is the kinematic viscosity.
The quantities $l$ and $u$ may differ essentially from 
the size of the streamline body and mean velocity around, respectively. 
Therefore for different flow systems 
the critical value of $Re$ can be very different and even approaches infinity, 
for example, in pipe flows~\cite{TL:2008}. 
In the paradigmatic setting of a uniform flow of velocity $u$ 
flowing past a cylinder of diameter $l$, 
when $Re$ is of the order of tens, the flow is regular. 
For $Re$ in the hundreds, 
von K\'{a}rm\'{a}n vortex street forms behind the cylinder~\cite{VanDyke:book}, 
breaking certain symmetries of the system. 
Fully developed turbulence, in which the broken symmetries are restored, 
occurs at very high Reynolds number, typically in the thousands. 
In situations where turbulence occurs at high Reynolds numbers, 
its study may be challenging, both experimentally where flow systems of 
enormous size and/or high velocity are required and computationally 
where unconventionally high resolution in the numerical integration of the 
Navier-Stokes equation is needed. It is thus desirable that 
fluid turbulence can emerge in physical flows of relatively low Reynolds numbers. 
Meanwhile the `way' that turbulence arises can be very different!
Consider the three-dimensional Navier-Stokes equations, 
mainly four routes were studied so far~\cite{GB:1980, FZ:1992}. 
These present special characteristics and states with 
(i) quasi-periodicity (with two frequencies on two tori) and phase locking, 
(ii) subharmonic (period doubling) bifurcations,
(iii) three frequencies (on two- and/or three tori), 
and (iv) intermittent noise. 

Thus it is known that fully developed turbulence may occur 
after a set of bifurcations~\cite{GS:1975}, 
as in a circular Couette flow with rotating inner cylinder. 
Moreover, transition to fully developed turbulence may depend 
not only on the Reynolds number, 
but on the particular characteristics of the flow evolutions. 
For example, in Ref.~\cite{GS:2000}, 
it was observed experimentally that 
the flow of a sufficiently elastic polymer solutions 
can become irregular even at low velocity, 
high viscosity and in a small tank that 
corresponds to very small Reynolds number (on the order of unity). 
This flow has all the main features of fully developed turbulence: 
a broad range of spatial and temporal scales. 
In general the understanding and ``turbulence control"~\cite{LB:1998,CMK:1994}
potentially has a large impact on society in particular in economical interests. 
The energy dissipation of turbulent flows is much larger than that of laminar ones. 
Consequently it is more costly to transport fluid through a vessel or to propel a vehicle 
if the flow is turbulent. 
For example, in oil pipelines the pressures required to pump the fluid is 
typically up to thirty times larger than would be necessary 
if the flow could be held laminar. 
Thus to keep the flow laminar until higher Reynolds numbers nowadays 
it is common to add polymers into the flow in oil pipelines.
Similar one could think to add ferrofluids in circular closed systems 
to keep the flow laminar due to an applied magnetic field.
Studying transition to turbulence in non-Newton fluid systems 
continues to be an interesting topic.

In this paper, 
we investigate the Taylor-Couette flow~\cite{Taylor:1923} in finite systems 
(e.g., aspect ratio $\Gamma=20$) 
where a rotating {\em ferrofluid}~\cite{Rosensweig:book} is confined by axial end walls, 
i.e., non-rotating lids, in the presence of an external magnetic field. 
The classic Taylor-Couette flow of non-ferrofluid has been 
a computational~\cite{CI:1994,AH:2010} and experimental~\cite{DS:1985,ALS:1986,Tagg:1994}
paradigm to investigate a variety of nonlinear and complex dynamical phenomena, 
including the transition to turbulence at high Reynolds numbers~\cite{Coles:1965}. 
The corresponding ferrofluid system we study consists of two independently rotating, 
concentric cylinders with viscous ferrofluid filled in between, 
which has embedded within itself artificially dispersed magnetized nanoparticles. 
In the absence of the external magnetic field, 
the magnetic moments of the nanoparticles are randomly oriented, 
leading to zero net magnetization for the entire fluid. 
In this case, 
the magnetized nanoparticles have little effect on the physical properties of the fluid 
such as density and viscosity. 
However, when a transverse magnetic field is applied, 
the physical properties of the fluid can be significantly modified~\cite{Rosensweig:book,Shliomis:1972}, 
leading to drastic changes in the underlying hydrodynamics. 
For example, for systems of rotating ferrofluid~\cite{Rosensweig:book}, 
an external magnetic field can stabilize regular dynamical states~\cite{AHLL:2010,RO:2011,ALD:2012,ALD:2013} 
and induce dramatic changes in the flow topology~\cite{AHLL:2010,RO:2011,HSS:2003}. 
(In fact, the effects of magnetic field are particularly important for 
geophysical flows~\cite{FL:2004,SL:2005,WJH:2012,TZL:2012}).
In general, the magnetic field can be used effectively as 
a control or bifurcation parameter of the system, 
whose change can lead to characteristically distinct types of 
hydrodynamical behaviors~\cite{AHLL:2010,RO:2011,ALD:2012,ALD:2013}.
In this regard, transition to turbulence in magnetohydrodynamical (MHD) flows 
with current-driven instabilities of helical fields 
has been investigated~\cite{GRH:2011}. 
There is also a large body of literature on MHD dynamics in 
Taylor-Couette flows~~\cite{SGGHPRS:2009,HTR:2010,SSGWG:2012,RSGESGJ:2012}.
Existing works on rotating ferrofluids~\cite{VNS:1986,Niklas:1987,Hart:2002,AHLL:2010,RO:2011,ALD:2012}, 
however, are mostly concerned with steady {\em time-independent} flows. 
Time-dependent ferrofluid flows have been investigated only recently but 
in the non-turbulent regime~\cite{Altmeyer:2013}.

The interaction between ferrofluid and magnetic field leads to 
additional terms in the Navier-Stokes equation~\cite{AHLL:2010,ALD:2013,Niklas:1987}. 
Our extensive computations reveals 
a sequence of bifurcations leading to time-dependent flow solutions 
such as standing waves with periodic or quasiperiodic oscillations, and turbulence. 
Surprisingly, we find that turbulence can occur for Reynolds numbers 
at least one order of magnitude smaller than 
those required for turbulence to arise in conventional fluids. 
The occurrence of turbulence is ascertained 
by a bifurcation analysis and by examining the characteristics of the physical 
quantities such as the energy, the wave number, and the angular momentum. 
We also find that the onset of turbulence can be determined accurately, 
in contrast to classical fluid turbulence 
where such a determination is typically qualitative and involves 
a high degree of uncertainty~\cite{Frisch:book}. 
Our findings have the following implications: 
\begin{itemize}
\item 
Ferrofluids under magnetic field is a new paradigm for investigating 
turbulence, especially experimentally where the study can be greatly facilitated 
due to the dramatic relaxation in the Reynolds-number requirement. 
\item The critical magnetic-field strength for the onset of turbulence 
can be pinned down precisely, possibly leading to deeper insights into
the physical and dynamical origins of the transition. 
\item Turbulence can be controlled externally, e.g., by an external magnetic field.
\end{itemize}

\noindent
{\large\bf Results}
\vspace*{0.1in}
\paragraph*{Ferrohydrodynamical equation of motion.}
Consider a Taylor-Couette system consisting of two concentric,
independently rotating cylinders with an incompressible, isothermal, homogeneous, 
mono-dispersed ferrofluid of 
kinematic viscosity $\nu$ and density $\rho$ within the annular gap. 
The inner and outer cylinders of radii $R_1$ and $R_2$ rotate 
at angular speeds $\omega_1$ and $\omega_2$, respectively. 
The top and bottom end-walls~\cite{HF:2004} are stationary and 
are at distance $\Gamma (R_2-R_1)$ apart, 
where $\Gamma$ is the non-dimensional aspect ratio. 
The system can be described using a cylindrical polar coordinate system $(r,\theta,z)$ 
with velocity field $(u_r,u_{\theta},u_z)$ and the corresponding vorticity 
$\nabla \times \bm{u} = (\xi,\eta,\zeta)$. 
In our preliminary study 
we set the radius ratio of the cylinders and the parameter $\Gamma$ to typical values 
in experiments, e.g., $R_1/R_2=0.5$ and $\Gamma=20$.
An external, homogeneous magnetic field of strength $H_x$ 
is applied in the transverse $x$-direction ($x=r\cos{\theta}$). 
The gap-width $d=R_2-R_1$ can be chosen as the length scale and 
the diffusion time $d^2/\nu$ can serve as the time scale. 
The pressure can be normalized by $\rho\nu^2/d^2$, and the magnetic field ${\bf H}$ 
and the magnetization ${\bf M}$ by $\sqrt{\rho/\mu_0} \nu/d$, 
where $\mu_0$ is the magnetic permeability of free space. 
We then obtain the following non-dimensionalized equations 
governing the flow dynamics~\cite{ALD:2013,ML:2001}:
\begin{eqnarray}
\nonumber
(\partial_t + {\bf u \cdot \nabla}) {\bf u} - \nabla^2 {\bf u}+ \nabla p 
&=& ({\bf M}\cdot \nabla) {\bf H} + 
\frac{1}{2} \nabla \times ({\bf M}\times {\bf H}), \\
\label{EQ:nast}
\nabla \cdot {\bf u} &=& 0.
\end{eqnarray}
The boundary conditions on the cylinders are 
${\bf u}(r_1,\theta,z)=(0,Re_1,0)$ and ${\bf u}(r_2,\theta,z)=(0,Re_2,0)$, 
where the inner and outer Reynolds numbers are 
$Re_1=\omega_1 r_1 d/\nu$ and $Re_2=\omega_2 r_2 d/\nu$, respectively, and 
$r_1=R_1/(R_2-R_1)$ and $r_2=R_2/(R_2-R_1)$ are 
the non-dimensionalized inner and outer cylinder radii, respectively.
To be concrete, we fix the Reynolds numbers at $Re_1=100$ and $Re_2=-150$
so that the rotation ratio $\beta = Re_2/Re_1$ of the cylinders is $-3/2$.

Equation~\eqref{EQ:nast} is to be solved together with an equation that
describes the magnetization of the ferrofluid. 
Using the equilibrium magnetization of an unperturbed state 
where homogeneously magnetized ferrofluid is at rest and 
the mean magnetic moments is orientated in the direction of the magnetic field, 
we obtain ${\bf M^\text{eq}} = \chi {\bf H}$. 
The magnetic susceptibility of the ferrofluid, $\chi$, 
can be determined by Langevin's formula~\cite{Langevin:1905}. 
The ferrofluids considered correspond to APG933~\cite{EMWKL:2000} with $\chi=0.9$. 
The near-equilibrium approximations of Niklas~\cite{Niklas:1987,NML:1989} are
small $||{\bf M} - {\bf M}^\text{eq}||$ and 
small relaxation time $|\nabla \times {\bf u}| \tau \ll 1$, 
where $\tau$ is the magnetic relaxation time. 
Using these approximations, 
Altmeyer et al.~\cite{ALD:2013} obtained the following equation:
\begin{equation}
\label{EQ:niklas}
{\bf M} - {\bf M}^\text{eq} = c^2_N 
\left( \nabla \times {\bf u} \times {\bf H}/2 
+ \lambda_2  {\mathbb S} {\bf H} \right),   
\end{equation}
where
\begin{equation}
c^2_N =\tau / \left(\displaystyle 1/\chi + \displaystyle \tau
\mu_0 H^2 / 6\mu\Phi\right)
\end{equation}
is the Niklas coefficient, $\mu$ is the dynamic viscosity, 
$\mu_0$ the vacuum viscosity, 
$\Phi$ is the volume fraction of the magnetic material, 
${\mathbb S}$ is the symmetric component of the velocity gradient tensor, and 
$\lambda_2$ is the material-dependent transport coefficient~\cite{ML:2001}. 
We choose $\lambda_2 = 2$, which corresponds to strong particle-particle interaction 
and chain formation of the ferrofluid~\cite{ML:2001}.

Using Eq.~\eqref{EQ:niklas}, 
we can eliminate the magnetization from Eq.~\eqref{EQ:nast} to obtain 
the following ferrohydrodynamical equation of motion~\cite{ML:2001,ALD:2013}:
\begin{eqnarray} \label{FHD}  
&   &  (\partial_{t} + {\bf u}\cdot \nabla ) {\bf u} - \nabla^{2} 
       {\bf u} + \nabla p_M \\ \nonumber
& = & - \frac{ c^2_N}{2} [{\bf H} \nabla \cdot 
       ({\bf F} + \lambda_2 {\mathbb S} {\bf H}) + {\bf H} \times 
       \nabla \times ({\bf F} + \lambda_2 {\mathbb S} {\bf H})],
\end{eqnarray}
where ${\bf F} = (\nabla \times {\bf u}/2) \times {\bf H}$ and $p_M$ is 
the dynamic pressure incorporating all magnetic terms 
that can be written as gradients. 
To leading-order approximation, the internal magnetic field in the ferrofluid 
can be regarded as being equal to the externally imposed field~\cite{ALD:2012}, 
which is reasonable for obtaining dynamical solutions of the magnetically driven fluid motion.
Equation~\eqref{FHD} can then be simplified as
\begin{eqnarray} \label{EQ:NSE_without_M_special}
&   & (\partial_t + {\bf u \cdot \nabla}) {\bf u} - \nabla^{2}
          {\bf u} + \nabla p_M 
 =  s_x^2 \left\lbrace \nabla^2 {\bf u} - 2 \lambda_2 
         \left[ \nabla \cdot ({\mathbb S} {\bf H}) \right] \right. \\ \nonumber 
& & \ \ \ \ \ \ \ - {\bf H} \times \left[ 
      \nabla \times (\nabla \times {\bf u} \times {\bf H}/2)  
 - {\bf H} \times (\nabla^2 {\bf u}) \right. 
\left. \left. + \lambda_2 \nabla \times ({\mathbb S} {\bf H}) 
    \right] \right\rbrace .
\end{eqnarray}
This way, the effect of the magnetic field and the magnetic properties of the ferrofluid 
on the velocity field can be characterized by a single parameter, 
the magnetic field or the Niklas parameter,
\begin{equation}
 s_x = \cfrac{2(2+\chi)H_x c_N}{(2+\chi)^2-\chi^2\eta^2}.
\end{equation}
\begin{figure}
\includegraphics[width=0.8\linewidth]{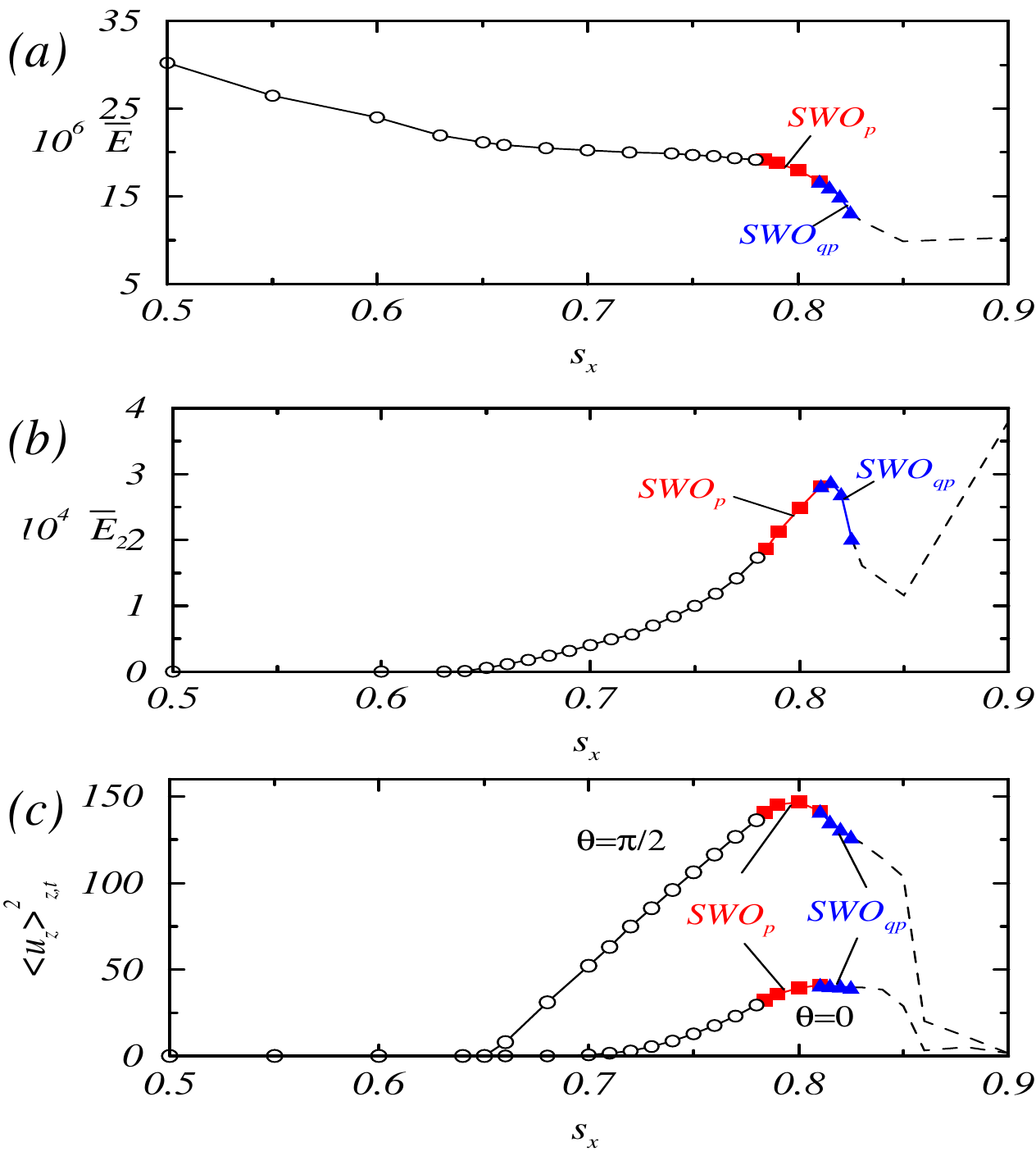}
\caption{Bifurcations with Niklas parameter $s_x$: 
(a) time-averaged modal kinetic energy, 
(b) its $m = 2$ contribution, and 
(c) spatiotemporally averaged axial flow field at midgap 
for $\theta=0$ and $\theta=\pi/2$ in respect of the applied magnetic field.
Open and filled symbols are for 
steady-state and time-dependent solutions, respectively.}
\label{fig:bif_sx}
\end{figure} 

\begin{figure}[h]
\includegraphics[width=0.8\linewidth]{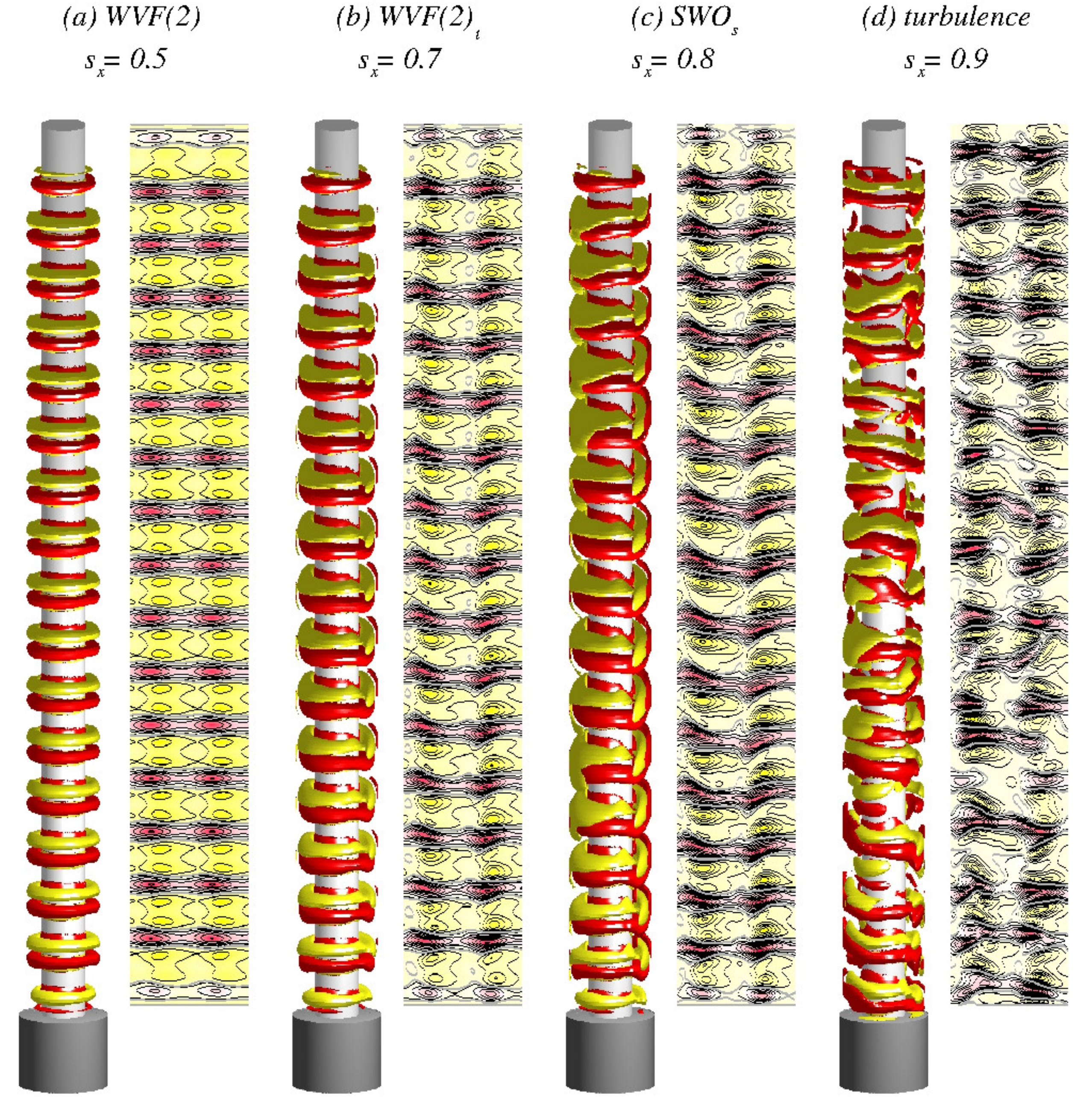}
\caption{(a-d) For four values of $s_x$ corresponding to WVF$_2$, WVF$_t$, 
SWO$_p$, and turbulence regimes, respectively, isosurfaces of azimuthal 
vorticity $\eta=\partial_z u_r - \partial_r u_z$ and contours of the radial 
velocity $u_r(\theta,z)$ on an unrolled cylindrical surface in the annulus 
at midgap. Red (dark gray) and yellow (light gray) colors denote $\eta=\pm 100$ 
for isosurfaces, and inflow and outflow for contour plots, respectively.
While the pattern in (a,b) are stationary, the ones in (c,d)
are snapshots due to time dependence of the corresponding flow.}  
\label{fig:phi-z}
\end{figure}

\paragraph*{Transition to turbulence in ferrofluids.}
We first present a sequence of bifurcations with the magnetic parameter $s_x$, 
eventually leading to turbulence. 
For $s_x \ne 0$, the rotating ferrofluid flow 
is intrinsically three-dimensional~\cite{AHLL:2010,RO:2011,ALD:2012,ALD:2013} 
with increased complexity as $s_x$ is increased from zero. 
Figures~\ref{fig:bif_sx}(a-c) show, respectively, three key quantities versus $s_x$: 
the time-averaged modal kinetic energy defined as 
\begin{equation}
\sum_{m} \overline{E}_m = \left\langle 
\int_0^{2\pi} \int_{-\Gamma/2}^{\Gamma/2} \int_{r_i}^{r_o} 
{\bf u}_m {\bf u}^*_m r \textrm{d}r \textrm{d}z \textrm{d}\theta
\right\rangle_t
\end{equation}
its $m = 2$ contribution $\overline{E}_2$, 
and the spatiotemporally averaged axial flow field 
$\langle u_z\rangle_{z,t}^2$ at midgap for $\theta=0$ (along the magnetic field) 
and $\theta=\pi/2$ (perpendicular to the field). 
We see that, for $s_x < 0.784$, the flow is {\em time-independent}. 
In particular, for $0 < s_x < s_x^{Wt} \approx 0.642$, 
the flow patterns show wavy vortices with a two-fold symmetry~\cite{AHLL:2010}, 
denoted as WVF(2), 
which appears in Fig.~\ref{fig:phi-z}(a) as a ``two-belly'' structure. 
As $s_x$ is increased through $s_x^{Wt}$, 
this symmetry is broken and a somewhat tilting pattern in the wavy-vortex structure 
emerges [denoted as WVF(2)$_t$], as shown in Fig.~\ref{fig:phi-z}(b). 
The intuitive reason is that 
the magnetic force is downward for fluid flow near the annulus and 
upward where the flow exits, 
resulting in a split in $\langle u_z\rangle_{z,t}^2$ for $\theta=0$ and $\pi/2$. 
As $s_x$ is increased through the critical point $s_x^p \approx 0.784$,
the flow becomes {\em time-dependent}. 
For $s_x \agt s_x^p$, the flow is time periodic (limit cycle), 
corresponding to standing waves 
with axial oscillations of characteristic frequency $\omega_1$, 
which is denoted as SWO$_p$. 
Near the onset of SWO$_p$, 
the kinetic energy values $\overline{E}$ and $\overline{E}_2$ are unaffected, 
as can be seen from Figs.~\ref{fig:bif_sx}(a) and \ref{fig:bif_sx}(b), respectively. 
For $s_x \agt s_x^{qp} \approx 0.81$, 
the periodic solution becomes unstable 
due to the emergence of the second incommensurate frequency 
$\omega_2$ ($\approx \omega_1/5$) associated with defect propagation 
through the annulus in the axial direction, 
leading to a transition to quasiperiodic flow (denoted as SWO$_{qp}$).
[See movie files movie1.avi, movie2.avi, movie3.avi, and movie4.avi
in Supplementary Materials (SMs),
where the additional frequency can be identified visually 
in the defect starting from the top of the oscillating region, 
propagating downward toward the bottom, and vanishing there.] 
While $\overline{E}$ continues to decrease through this transition, 
both $\overline{E}_2$ and $\langle u_z\rangle_{z,t}^2$ reach 
their respective maxima at the transition point. 
As $s_x$ is increased further, the flow becomes more complex. 
Onset of turbulence occurs for $s_x \approx s_x^{tu} \approx 0.825$, 
where globally the flow starts to rotate in the azimuthal direction, 
as shown in Fig.~\ref{fig:phi-z}(d)
(see also movies movie5.avi, movie6.avi, and movie7.avi in SMs).
The observed route to turbulence coincides completely with 
that established previously~\cite{GS:1975,RT:1971} 
for conventional fluid at high Reynolds numbers. 

\begin{figure}
\includegraphics[width=0.8\linewidth]{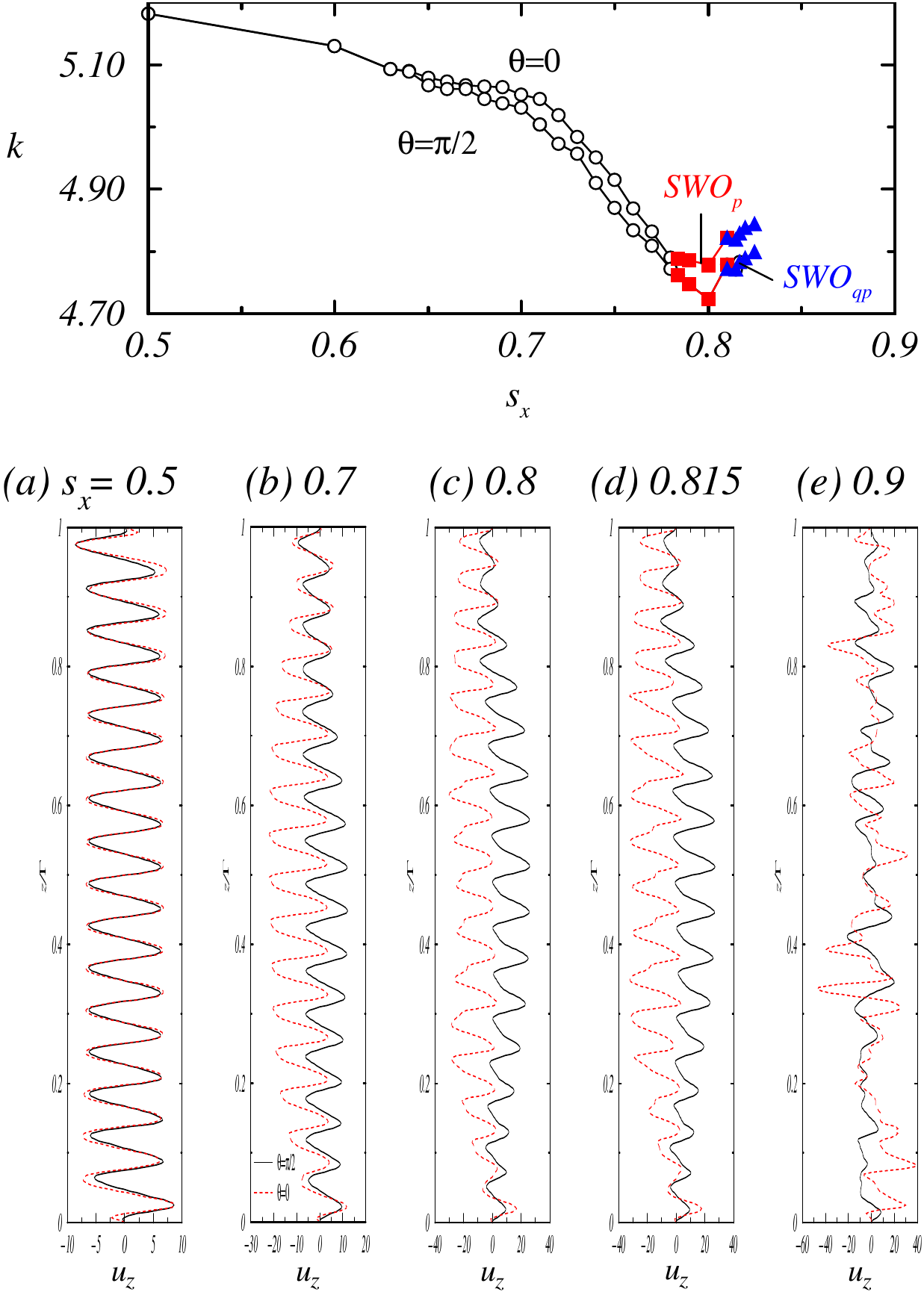}
\caption{Top panel: variation with $s_x$ of the axial wave number $k$
in the directions along ($\theta=0$) and perpendicular to ($\theta=\pi/2$) 
the magnetic field. (a-e) Snapshots of axial velocity $u_z$ for $\theta=0$ 
(dashed lines) and $\theta=\pi/2$ (solid lines) in the annulus at the midgap 
location for five different values of $s_x$, where (c-e) correspond to periodic, 
quasiperiodic, and turbulent flows, respectively. The presented patterns
are for $m=0$ so that we can identify the largest contribution in the
axial Fourier spectrum of $(u_z)_0(z,t)$. In principle one can also identify
$k$ from the axial profiles in the figure that gives the axial
wavelength $\lambda$ and consequently the wavenumber $k=2\pi/\lambda$.
See also movie files 
movie8.avi, movie9.avi, movie2.avi, movie4.avi and movie7.avi in SMs.
}
\label{fig:wavenumber}
\end{figure}

\paragraph*{Wave structure of the flows.}
We next analyze the wave structure of the flows 
to probe into the transition to turbulence in the low Reynolds-number regime. 
To get numerical solutions of the ferrohydrodynamical system,
we solve the ferrohydrodynamical system by combining a finite-difference, 
time-explicit method of second order in $(r,z)$ 
with Fourier spectral decomposition in $\theta$ (See {\bf Method}). 
The variables can be written as 
\begin{equation}
f(r,\theta,z,t)=\sum_{m=-m_{\max}}^{m_{\max}} f_m(r,z,t)\,e^{im\theta},
\end{equation}
where $f$ denotes any one of $\{u_r,u_\theta,u_z,p\}$. 
We then perform an axial Fourier analysis~\cite{AHLL:2010} 
of the mode amplitudes of the radial velocity field $(u_r)_m(z,t)$ at midgap 
to obtain the behavior of the wave number $k$, as shown in Fig.~\ref{fig:wavenumber}. 
Prior to the onset of turbulence, 
the number of vortices in the bulk fluid is invariant, 
regardless of whether the flow is time-independent or time-dependent. 
However, due to the symmetry breaking induced by the magnetic field, 
the spatial structures of the flows can differ significantly. 
The top panel of Fig.~\ref{fig:wavenumber} shows that 
the axial wave number $k$ begins to split into two branches at $s_x = s_x^{Wt}$, 
where $k$ is enhanced and reduced as 
the field enters into and exits perpendicularly from the annulus, respectively, 
and there are dramatic changes in the flow profiles 
in which the dynamics tend to focus on the $\theta \approx 0$ region. 
As $s_x$ is increased further, 
$k$ in the bulk decreases with nearly constant difference 
between the cases of $\theta=0$ and $\theta=\pi/2$. 
However, the kinetic energies and the axial wave numbers are nearly unaffected 
when the flow becomes time-dependent as $s_x$ is increased through $s_x^p$. 
As the quasiperiodic regime is reached, 
the energies and the axial wave numbers begin to change where, 
as shown in Fig.~\ref{fig:wavenumber}, 
the wave numbers for $\theta=0$ and $\theta=\pi/2$ are minimized at $s_x$ about 0.8. 
For both the periodic and quasiperiodic regimes, 
there is little variation in $k$ and 
only the vortex pair surrounding the oscillatory region expands and shrinks periodically. 
Especially, in the periodic regime 
the flow profile at $\theta=\pi/2$ exhibits little dependence on $s_x$ 
except for an increasing amplitude, 
but in the quasiperiodic regime the flow shows a strong dependence on $s_x$, 
due to the emergence of the second frequency. 
As $s_x^{tu}$ is passed, 
the flow starts to {\em rotate} and the split in the axial wave numbers 
in the directions parallel and perpendicular to the magnetic field vanishes. 
In fact, as turbulence sets in no distinct wave numbers can be identified, 
as shown in Figs.~\ref{fig:phi-z} and \ref{fig:wavenumber}(e) 
(see also movie file movie7.avi in SMs).

\begin{figure}[h]
\includegraphics[width=0.8\linewidth]{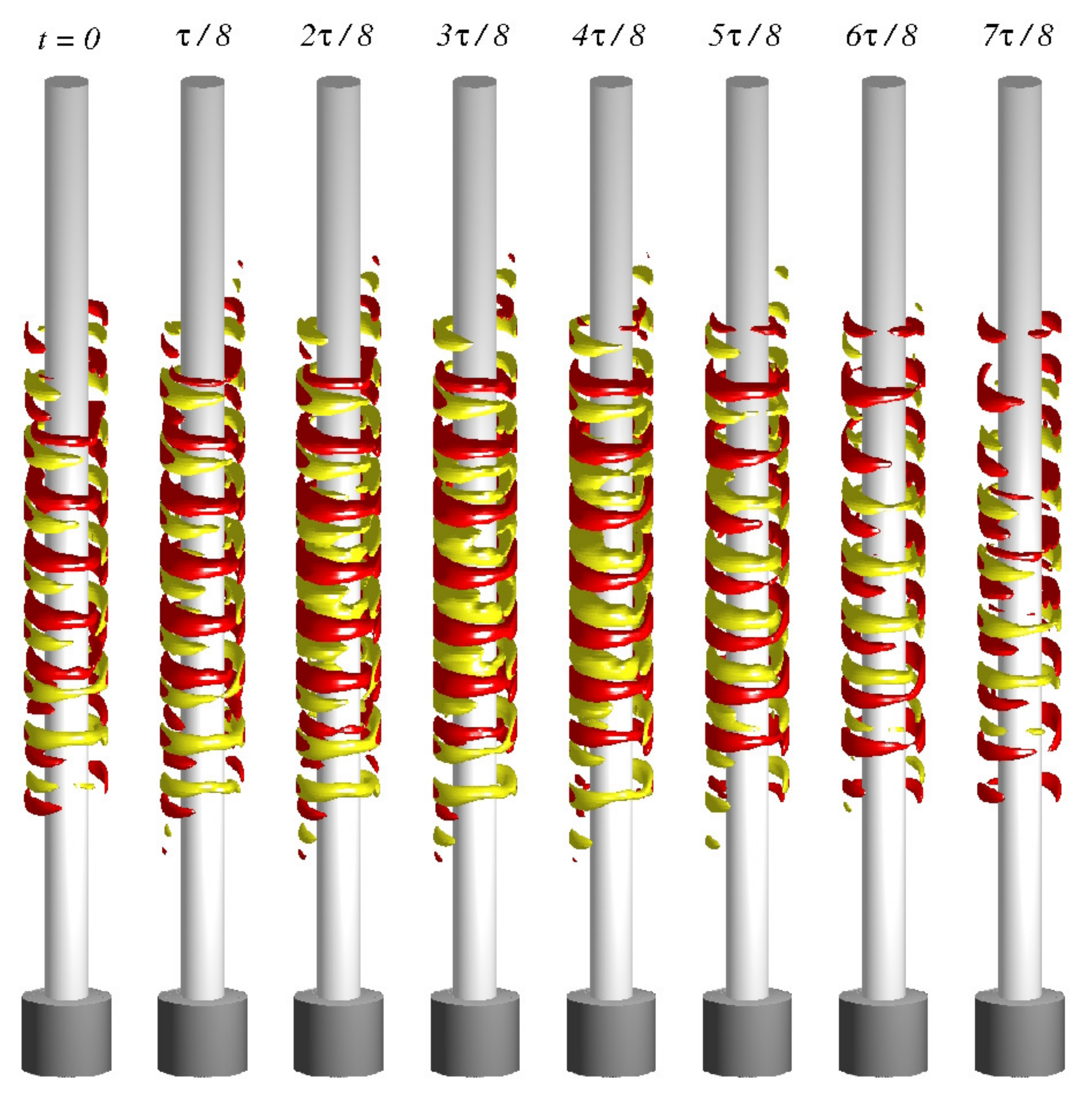}
\caption{For $s_x = 0.8$ (periodic regime), isosurfaces of the relative angular 
momentum $ru_\theta - \int_0^\tau ru_\theta dt$ at eight time instants in one 
period of oscillation, where $\tau \approx 0.058$ and the isolevels are 
$ru_\theta=\pm5$ 
(see also movie files movie10.avi, movie11.avi, and movie12.avi in SMs).
}
\label{fig:RBC_M-wTVF}
\end{figure} 

\paragraph*{Behavior of the angular momentum.}
We now examine the behavior of the angular momentum 
in relation with the flow bifurcation sequence and transition to turbulence. 
Figure~\ref{fig:RBC_M-wTVF} shows, for flows in the periodic regime, 
the isosurfaces of differences in the angular momentum $ru_\theta$ 
between the full flow and its long-time averaged value, 
namely, $ru_\theta - \int_0^\tau ru_\theta dt$. 
Onset of the periodic regime can be identified 
by the occurrence of periodic oscillations (up and down) of 
a single outward directed jet of angular momentum $ru_\theta$ 
(see movie files movie2.avi, and movie1.avi in SMs).
From Fig.~\ref{fig:RBC_M-wTVF}, we see that, 
while the full flow pattern in the periodic regime exhibits little variation
(see movie files movie10.avi and movie11.avi in SMs),
there is relatively large variation in the behavior of the angular momentum 
(see movie12.avi in SMs). 
For example, 
there is a downward flow from $\theta=0$ and an upperward flow from $\theta=\pi$, 
with opposite angular-momentum values. 
In the middle of the finite system 
where the effects of the Ekman boundary layers are minimal, 
the difference in the angular momentum reaches maximum. 
As the system enters into the quasiperiodic regime, 
the appearance of the incommensurate frequency $\omega_2$ signifies 
a kind of defects in the propagation pattern of the flow. 
As the flow begins to rotate in the azimuthal direction, turbulence sets in.

\begin{figure}[h]
\includegraphics[width=0.65\linewidth]{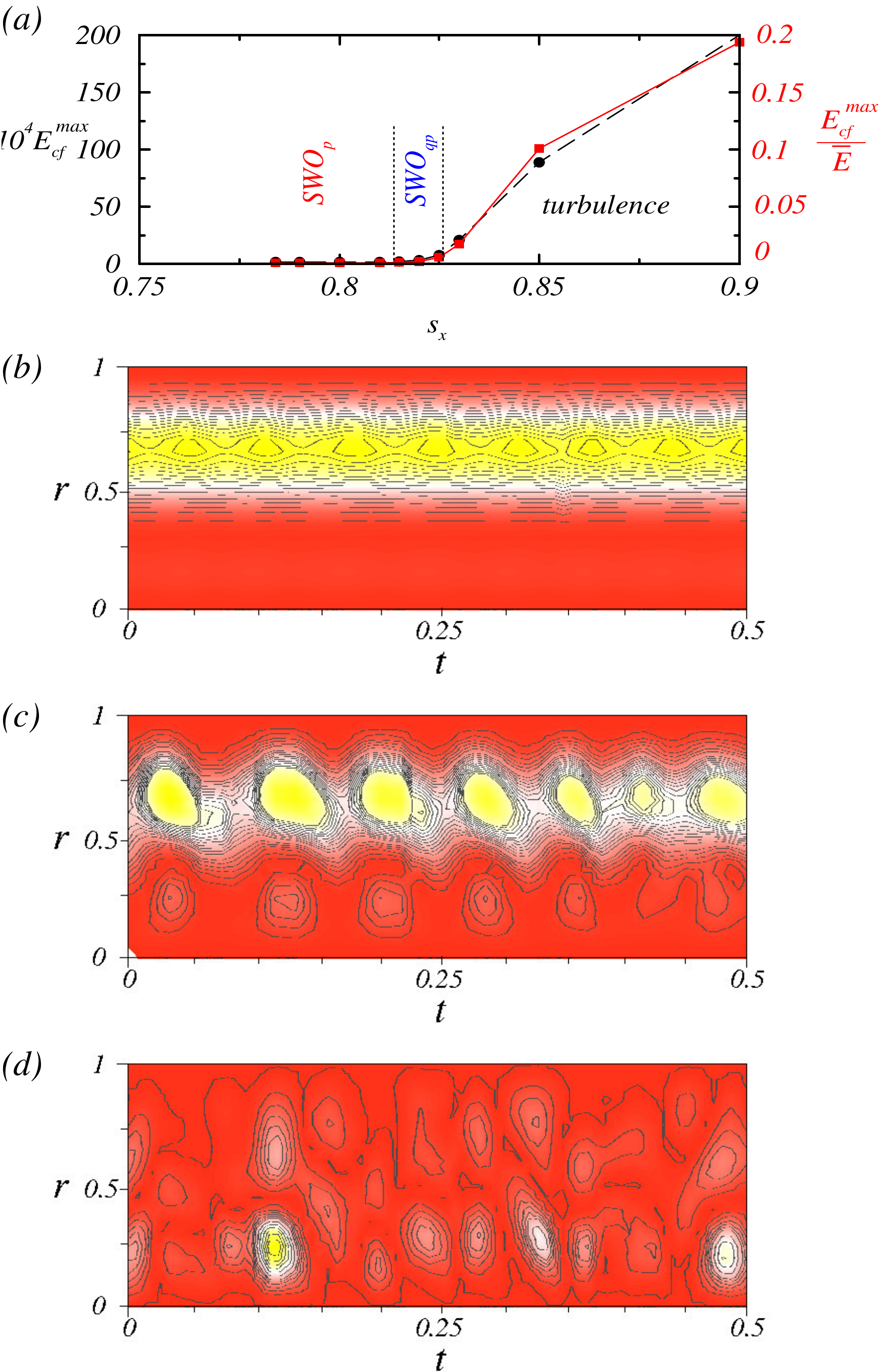}
\caption{(a) Unnormalized and normalized (by the total kinetic energy)
maximum cross-flow energy $E^{max}_{cf}$ versus the Niklas parameter 
$s_x$, where $E_{cf}(r,t) \equiv \langle u_r^2+u_z^2\rangle_{A(r)}$ is 
averaged over the surface of the concentric cylinder. (b-d) Spatiotemporal 
evolutions of $E_{cf}(r,t)$ for $s_x = 0.8$, $0.82$, and $0.9$, corresponding 
to periodic, quasiperiodic, and turbulent regimes, respectively. Red (yellow) 
color indicates high (low) energy values.}   
\label{fig:crossflow-energy}
\end{figure} 

\paragraph*{Low Reynolds number turbulence.}
An extremely challenging issue in the study of turbulence is 
precise determination of its onset as a system parameter, 
e.g., the Reynolds number, is changed. 
For the type of low-Reynolds number turbulence in ferrofluid uncovered in this Letter, 
this can actually be achieved. 
In particular, the transition to turbulence coincides with 
the onset of azimuthal rotation from the quasiperiodic flow. 
To illustrate this, we consider the cross-flow energy $E_{cf}(r,t)$, 
a commonly used indicator 
for the onset of turbulence in the Taylor-Couette system~\cite{BE:2013}, 
which is the instantaneous energy associated with 
the transverse velocity component at radial distance $r$. 
Small (large) values of $E_{cf}$ indicate laminar (turbulent) flows. 
We find that, the maximum value of $E_{cf}(r,t)$, denoted by $E^{max}_{cf}$, 
assumes near zero values for $s_x < s_x^{qp}$ 
but it increases dramatically as $s_x$ is increased through $s_x^{qp}$, 
as shown in Fig.~\ref{fig:crossflow-energy}(a). 
In this figure we also plot the scaled maximum cross-flow energy value,
$E^{max}_{cf}/\overline{E}$, versus $s_x$, and we observe that 
$E^{max}_{cf}$ and $E^{max}_{cf}/\overline{E}$ exhibit essentially the same behavior 
as the system passes through the onset of turbulence. 
Examples of the spatiotemporal evolution of $E_{cf}(r,t)$ 
are shown in Figs.~\ref{fig:crossflow-energy}(b,c,d) for periodic, 
quasiperiodic, and turbulent regimes, respectively. 
We observe regular patterns in the former two regimes, 
but no apparent patterns in the turbulent regime. \\
\\ \noindent
{\large\bf Discussion} \\
\\ \noindent 
To summarize, we have discovered that 
in the Taylor-Couette ferrofluid system driven by a magnetic field, 
where flow can exhibit axial oscillations 
but not rotations in the azimuthal direction in the regular regime, 
turbulence can generically arise and 
its onset can occur for low values of the Reynolds number. 
This is substantiated by 
extensive computations of the underlying ferrohydrodynamical equation 
through a systematic bifurcation analysis and 
characterization of behaviors of physical quantities. 
The implications, 
besides the surprising phenomenon of turbulence at very low Reynolds numbers 
and consequently facilitation of experimental study of turbulence, 
lie in the perspective of controlled generation of turbulence 
through variations of the external magnetic field, 
making it possible to locate the onset of turbulence with high precision. 
We expect these findings to have values for experimental 
as well as theoretical investigation of turbulence. \\
\\ \noindent
{\large\bf Method}
\vspace*{0.1in}
\paragraph*{Numerical scheme for ferrodynamical equation.}
System~\eqref{FHD} can be solved~\cite{AHLL:2010,ALD:2012,ALD:2013}
by combining a second-order finite-difference scheme in $(r,z)$ 
with Fourier spectral decomposition in $\theta$ and (explicit) time splitting. 
The variables can be written as
\begin{equation}
f(r,\theta,z,t)=\sum_{m=-m_{\max}}^{m_{\max}} f_m(r,z,t)\,e^{im\theta},
\end{equation}
where $f$ denotes one of $\{u_r,u_{\theta},u_z,p\}$. For the parameter regimes
considered in our preliminary study, 
the choice $m_{\max}=10$ provides adequate accuracy. 
We use uniform grids with spacing $\delta r = \delta z =0.05$
and time-steps $\delta t < 1/3800$. 
For diagnostic purposes, we also evaluate the complex mode amplitudes $f_{m,n}(r,t)$ 
obtained from the Fourier decomposition in the axial direction
$f_m(r,z,t)=\sum_{n} f_{m,n}(r,t)e^{inkz}$, where $k$ is the axial wavenumber.


\section*{Figure legends}

{\bf Figure 1}:
Bifurcations with Niklas parameter $s_x$: 
(a) time-averaged modal kinetic energy, 
(b) its $m = 2$ contribution, and 
(c) spatiotemporally averaged axial flow field at midgap 
for $\theta=0$ and $\theta=\pi/2$ in respect of the applied magnetic field. 
Open and filled symbols are for 
steady-state and time-dependent solutions, respectively.

{\bf Figure 2}:
(a-d) For four values of $s_x$ corresponding to WVF$_2$, WVF$_t$, 
SWO$_p$, and turbulence regimes, respectively, isosurfaces of azimuthal 
vorticity $\eta=\partial_z u_r - \partial_r u_z$ and contours of the radial 
velocity $u_r(\theta,z)$ on an unrolled cylindrical surface in the annulus 
at midgap. Red (dark gray) and yellow (light gray) colors denote $\eta=\pm 100$ 
for isosurfaces, and inflow and outflow for contour plots, respectively.
While the pattern in (a,b) are stationary, the ones in (c,d) are snapshots 
due to time dependence of the corresponding flow.

{\bf Figure 3}:
Top panel: variation with $s_x$ of the axial wave number $k$
in the directions along ($\theta=0$) and perpendicular to ($\theta=\pi/2$) 
the magnetic field. (a-e) Snapshots of axial velocity $u_z$ for $\theta=0$ 
(dashed lines) and $\theta=\pi/2$ (solid lines) in the annulus at the midgap 
location for five different values of $s_x$, where (c-e) correspond to periodic, 
quasiperiodic, and turbulent flows, respectively. The presented patterns
are for $m=0$ so that we can identify the largest contribution in the
axial Fourier spectrum of $(u_z)_0(z,t)$. In principle one can also identify
$k$ from the axial profiles in the figure that gives the axial
wavelength $\lambda$ and consequently the wavenumber $k=2\pi/\lambda$.
See also movie files 
movie8.avi, movie9.avi, movie2.avi, movie4.avi and movie7.avi in SMs.

{\bf Figure 4}:
For $s_x = 0.8$ (periodic regime), isosurfaces of the relative angular 
momentum $ru_\theta - \int_0^\tau ru_\theta dt$ at eight time instants in one 
period of oscillation, where $\tau \approx 0.058$ and the isolevels are 
$ru_\theta=\pm5$ 
(see also movie files movie10.avi, movie11.avi, and movie12.avi in SMs).

{\bf Figure 5}:
(a) Unnormalized and normalized (by the total kinetic energy)
maximum cross-flow energy $E^{max}_{cf}$ versus the Niklas parameter 
$s_x$, where $E_{cf}(r,t) \equiv \langle u_r^2+u_z^2\rangle_{A(r)}$ is 
averaged over the surface of the concentric cylinder. (b-d) Spatiotemporal 
evolutions of $E_{cf}(r,t)$ for $s_x = 0.8$, $0.82$, and $0.9$, corresponding 
to periodic, quasiperiodic, and turbulent regimes, respectively. Red (yellow) 
color indicates high (low) energy values.

\section*{Acknowledgement}
Y.D. was supported by Basic Science Research Program 
of the Ministry of Education, Science and Technology under 
Grant No.~NRF-2013R1A1A2010067.
Y.C.L. was supported by AFOSR under Grant No.~FA9550-12-1-0095.

\section*{Author contributions}
S.A., Y.D. and Y.C.L. devised the research project.
S.A. performed numerical simulations. 
S.A., Y.D. and Y.C.L. analyzed the results. 
S.A., Y.D. and Y.C.L. wrote the paper.

\section*{Additional information}


{\bf Competing financial interests}:
The authors declare no competing financial interests.

\end{document}